\documentclass[
  journal=pasa,
  manuscript=BGCE-Student-Paper-Prize, 
  year=2024,
  volume=37,
]{cup-journal}

\usepackage{amsfonts,amsmath,amssymb}

\usepackage{microtype,siunitx,booktabs}
\usepackage{listings}
\usepackage{pgf}
\usepackage{graphicx}
\usepackage{subcaption}
\usepackage{hyperref}
\usepackage{cleveref}
\usepackage{caption}
\usepackage{xcolor}
\usepackage{multicol}
\usepackage{algorithm}
\usepackage{algpseudocode}

\usepackage{times}

\usepackage{bm}
\usepackage{mathrsfs}
\usepackage{verbatim}
\usepackage{listings}
\usepackage{algpseudocode}
\usepackage{placeins}

\setcitestyle{numbers,square}
\definecolor{codegreen}{rgb}{0,0.6,0}
\definecolor{codegray}{rgb}{0.5,0.5,0.5}
\definecolor{codepurple}{rgb}{0.58,0,0.82}
\definecolor{backcolour}{rgb}{0.95,0.95,0.95}
\hypersetup{
	unicode=true,          
	pdftoolbar=true,        
	pdfmenubar=true,        
	pdffitwindow=false,     
	pdfstartview={FitH},    
	pdftitle={BGCE2024---Timothy Stokes},    
	pdfauthor={Timothy Stokes},     
	pdfsubject={Subject},   
	pdfcreator={Timothy Stokes},   
	pdfproducer={Timothy Stokes}, 
	pdfkeywords={}, 
	pdfnewwindow=true,      
	colorlinks=true,       
	linkcolor=orange,          
	citecolor=brown,        
	filecolor=brown,      
	urlcolor=brown           
}

\usepackage[final,commandnameprefix=ifneeded]{changes}
 
\definechangesauthor[name=TW, color=olive]{TW}
\definechangesauthor[name=TS, color=blue]{TS}

\title{Fast Higher-Order Interpolation and Restriction in ExaHyPE Avoiding Non-physical Reflections}

\author{Timothy Stokes}
\affiliation{Department of Computer Science, Durham University, United Kingdom}
\email[Timothy Stokes]{timothy.j.stokes@durham.ac.uk}

\author{Tobias Weinzierl}
\affiliation{Institute for Data Science and Department of Computer Science, Durham University, United Kingdom}

\author{Han Zhang}
\affiliation{Department of Computer Science, Durham University, United Kingdom}

\author{Baojiu Li}
\affiliation{Institute for Computational Cosmology, Durham University, United Kingdom}

\vspace{-1cm}

\begin{document}

\begin{abstract}
 Wave equations help us to understand phenomena ranging from earthquakes to tsunamis. These phenomena materialise over very large scales. It would be computationally infeasible to track them over a regular mesh. Instead, since the phenomena are localised, adaptive mesh refinement (AMR) can be used to construct meshes with a higher resolution close to the regions of interest. ExaHyPE is a software engine created to solve wave problems using AMR, and we use it as baseline to construct our numerical relativity application called ExaGRyPE. To advance the mesh in time, we have to interpolate and restrict along resolution transitions in each and every time step. ExaHyPE's vanilla code version uses a d-linear tensor-product approach. In benchmarks of black hole spacetimes this performs slowly and leads to errors in conserved quantities near AMR boundaries. We therefore introduce a set of higher-order interpolation schemes where the derivatives are calculated at each coarse grid cell to approximate the enclosed fine cells. The resulting methods run faster than the tensor-product approach. Most importantly, when running the black hole simulation using the higher order methods the errors near the AMR boundaries are removed.
\vspace{-1cm}
\end{abstract}

\section{Introduction}
\label{sec:intro}

Wave equations describe many phenomena of importance such as earthquakes with their destructive impact or gravitational waves which further our understanding of the Universe. These problems are too complex to solve analytically, so a numerical method is required. The ExaHyPE code (An Exascale Hyperbolic PDE Engine), was written for this purpose \cite{ExaHyPE}. We use it as baseline to construct our numerical relativity solver ExaGRyPE \cite{ExaGRyPE} which solves the Einstein equations in second- or first-order conformal and covariant Z4 (CCZ4)\cite{CCZ4} formulation employing high-order finite differences plus finite volume schemes.

Many wave phenomena of interest such as dynamic black holes and their gravitational waves have to be simulated over a large domain, and they also require high accuracy, so their simulation over a globally regular mesh would be very computationally expensive. Since the areas of interest with a huge impact on the global solution, such as a black hole, are localised, Adaptive Mesh Refinement (AMR) is used. ExaHyPE is built on top of Peano \cite{Peano}, which utilises a spacetree to construct an adaptive Cartesian mesh subject to 2:1 balancing \cite{tree_balancing}. This creates a non-conformal mesh with adjacent coarse and fine cells. To update the fine cells, ghost fine cells must be interpolated from the coarse data, and similarly coarse cells must be reconstructed, i.e.~restricted from the fine data. 

Finite Differences lend themselves towards tensor-product formulations in which the interpolation and restriction are separated into individual matrix multiplications in each dimension. We also naturally might assume that a trilinear interpolation and averaging are sufficient as we switch to coarser meshes if and only if the solution is sufficiently smooth. However, when we follow this train of thought in an ExaGRyPE simulation of a stationary black hole, we observe that the Hamiltonian constraint is violated at the AMR boundaries, an error which grows with time, indicating the insufficiency of the existing schemes.

We therefore propose to interpolate or restrict in every dimension at once using one large, yet sparse matrix. 
A second- or third- order interpolation and restriction can then be constructed by performing a Taylor expansion about each cell to obtain the derivatives at each grid cell up to second- or third- order respectively. As we restrict and interpolate into ghost cell regions, our method mirrors concepts of overlapping domain decomposition, such as in the Chimera method \cite{Chimera}. Increasing the order of accuracy for the method of interpolation up to the order of the discretisation scheme is consistently shown to increase the solution's overall accuracy \cite{GeneralOversetGrids,soton434785}, so this is an active area of research. Each matrix-based implementation outperforms its tensor-product cousin, once we take the block sparsity into account and pick a compressed matrix storage format.

Section \ref{section:problem_formulation} introduces Peano's data structures which underly ExaHyPE and hence ExaGRyPE, and rephrases the problem as a linear algebra setup. Section \ref{section:methodology} explains the mathematical basis for its method before we cover implementation details (Section \ref{section:implementation}). The manuscript concludes with a runtime and accuracy analysis in Section \ref{section:results} followed by a short outlook.
\section{Interpolation and Restriction in Peano}
\label{section:problem_formulation}

We employ a spacetree to construct an adaptive Cartesian meta mesh in Peano. This tree is constructed by starting with a single cube which spans the whole domain, and then is subdivided into three parts along each dimension. Where there is insufficient resolution, the resulting cube is recursively subdivided further. For applications involving travelling waves or shocks, we found it to be important that there are not sharp changes in the grid's resolution, so a 2:1 balancing method is used to ensure that unrefined cubes differ by at most one level of refinement \cite{tree_balancing}.

We call each cube that is not further refined in this scheme a patch, which hosts its own Cartesian mesh of dimension $p$ along each axis. Patches are the actual compute mesh. Per time step each patch is advanced in time separately. This requires each patch to have some knowledge of the adjacent patches. In ExaHyPE, a halo structure around each patch with a halo depth of $k\geq 1$ is used. $k\leq p$ depends upon the accuracy of the numerical scheme employed. For 4th order Finite Differences (FD), our standard solver for ExaGRyPE, we require $k=3$, while $k=1$ is sufficient for Finite Volumes (FV). 

The numerical schemes imply that we exclusively add halos along the patch faces. Diagonals across the edges or vertices are not required.

At the start of each time-step the halo values are updated using data from the face-connected neighbours, before we issue the actual time stepping compute kernel. For both FV and FD, we employ a cell-centred degree of freedom layout, i.e.~all 58 unknowns of the first-order CCZ4 formulations are associated with the centres of cells within the patches. As we equip all patches with halos, we effectively introduce an overlapping domain decomposition into patches.

\begin{figure}[ht]
    \centering
    \includegraphics[width=0.9\textwidth,bb=0 0 1510 982]{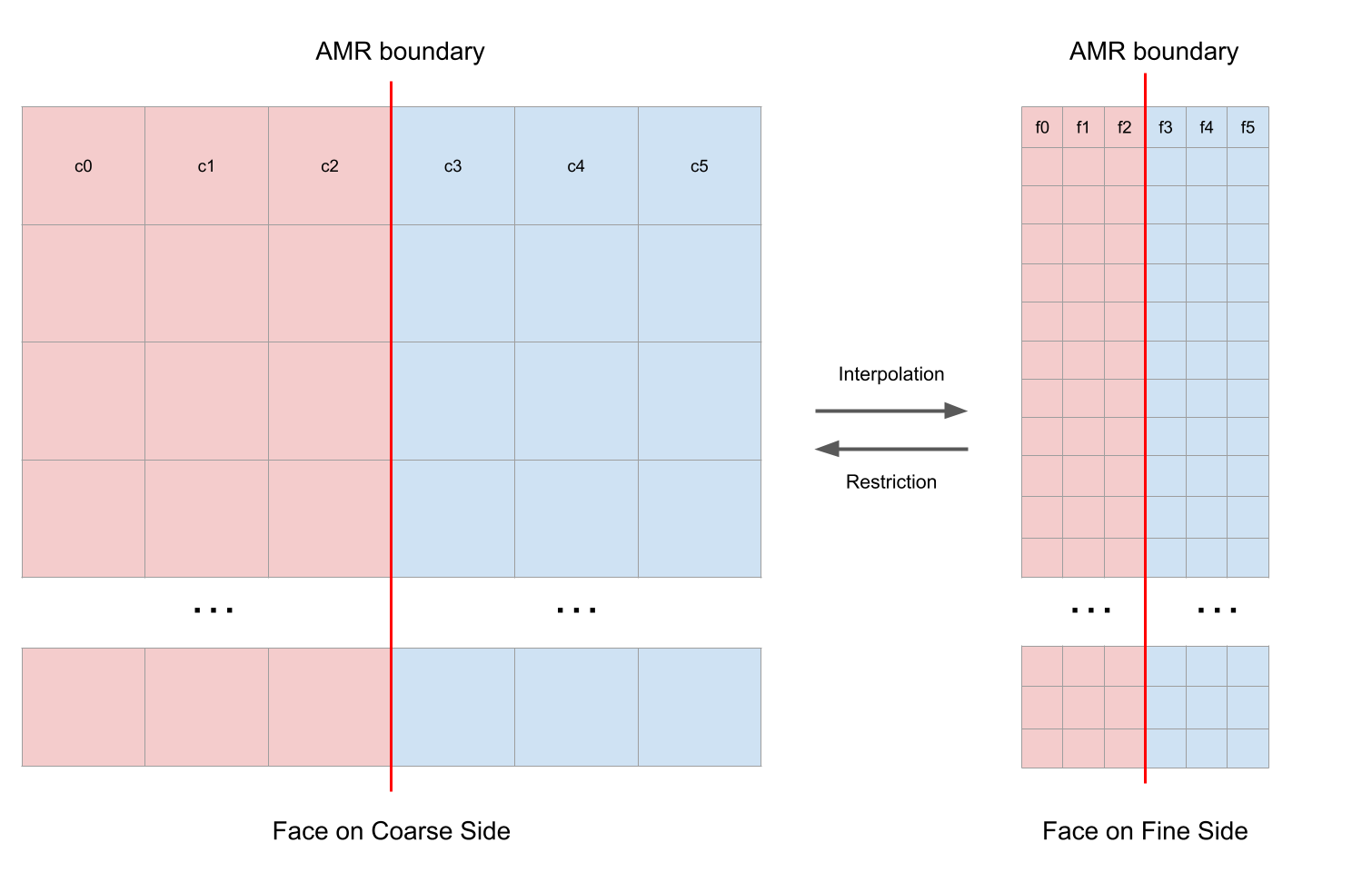}
    \caption{A diagram of the coarse and fine face structures at an AMR boundary \cite{ExaGRyPE}}
    \label{fig:interpolation}
\end{figure}

Where there is a change in resolution solely copying the 58 unknowns from the adjacent patch into the halo data structure is not sufficient, as there is no one-to-one relation between the coarse and fine cells. In these cases we have to interpolate into the halo of a patch from an adjacent coarser patch and in return restrict from the fine patches into the halos of adjacent coarser patches (Figure \ref{fig:interpolation}).

This can be reframed as a linear algebra problem, relating the coarse and fine face values. Notating the fine and coarse values as $\bm{Q^f}$ and $\bm{Q^c}$ respectively, we get
\begin{equation*}
  \bm{Q^f} = P\bm{Q^c}
  \qquad
  \text{or}
  \qquad
  \bm{Q^c} = R\bm{Q^f}
\end{equation*}
where $P$ is the interpolation matrix. The restriction matrix $R$ similarly relates the fine and coarse values.

 As we work with 2:1-balanced meshes, a patch has up to three unique adjacent halo layers in each direction which have to be interpolated.

ExaHyPE's code baseline uses linear interpolation or averaging  respectively, which yields a tensor product $P=P_xP_yP_z$ per face. The interpolation or restriction method of the reference face of a patch can be represented by two matrices $P_{\parallel}$ and $P_{\bot}$.
$P_{\parallel}$ interpolates tangentially to the AMR boundary and is applied twice subject to some permutation. The matrix $P_{\bot}$ interpolates normally to the boundary.

In three dimensions, a patch has 6 faces subject to interpolation.
We can reformulate $P$ and $R$ such that $P$ and $R$ are a single matrix per halo of the patch arrangement. For $P$ this matrix is of size $3p \cdot 3p \cdot k \times p \cdot p \cdot 2k = 9p^2k \times 2p^2k$. For $R$ it's of size $p \cdot p \cdot k \times 3p \cdot 3p \cdot 2k = p^2k \times 18p^2k$.

To fill a halo layer, we always take the halo layer in both directions around a separating face into account ($2k$).
As we usually only fill one halo layer of one patch or only restrict from one halo, we evaluate only a $1/9$ segment of $P$ or $R$ in one rush.
Factors of $3p$ and $9$ result from the three-partitioning in the spacetree.
For the unknowns associated with each degree of freedom, each quantity is interpolated/restricted independently subject to the same operator.
Both $P$ and $R$ are sparse.

Once reduced to a single matrix, there are six face configurations to consider.
Yet, we note that all of them result from one reference configuration, i.e.~we can rotate and mirror all geometric arrangements onto a reference face.
\section{Operator construction}
\label{section:methodology}

To construct higher-order interpolation operators, we identify per fine grid cell its closest coarse grid cousin.
As we work with three-partitioning, some fine grid cell centres coincide spatially with coarse centres.
We then employ a Taylor expansion starting from the centre of this coarse volume.

For Taylor, we need the derivatives in the coarse grid cell centres.
For the second order implementation we require $n=1+3+6=10$ coefficients, and $n=20$ in the third order case. 

To determine the derivatives, we fit polynomials through the coarse cell centres. Further details of which coarse cells we choose to fit these polynomials are in \ref{appendix:stencil}.

The polynomial fitting yields an $s \times n$ matrix $A$ per coarse grid cell, where $s$ is the stencil size, i.e.~the number of coarse grid cell centres entering the equation, and $n$ is the number of derivatives. The derivatives $\bm{x}$ can then be calculated from $A\bm{x}=\bm{Q^c}$ using standard matrix inversion methods \cite{MatrixComputations} and fed subsequently into the interpolation operator according to Taylor's formula.

The construction of a higher-order restriction is very similar. 
We identify per coarse cell the fine cell that is closest to its centre, then employ a Taylor expansion around this cell on the fine grid. 
After following the same process from here until the derivatives are obtained we calculate the coarse cell value, and that of any other coarse cells with the same nearest fine cell, using a Taylor expansion.
In the case where $k>1$ the closest fine cell lies directly on one of the coarse cell's centre, reducing the problem to a simple injection.
For the remaining coarse cells, an extrapolation is required.


\section{Implementation}
\label{section:implementation}

%
%
%
%
%

\subsection{Explicit matrix rotations}



After the collapsing of all interpolation operations along the faces of a patch, we are left with six $P$ and six $R$ matrices.
We realise that these are all permutations, i.e.~mirrored and rotated versions, of two reference matrices.
Indeed, we can pick one $P$ and one $R$ matrix and store exclusively this variant.
We choose the ones that correspond to an interpolation/restriction onto a halo layer with a normal along the x-axis.
Prior to the interpolation or restriction, all input data are copied into a temporary array.
They are projected onto a reference coordinate system.
We then apply $P$ or $R$ respectively before we copy the outcome back into the actual image halo.

Despite the additional memory allocations and moves, we find this solution, in the worst case, to be only 1.5\% slower than a version which stores all six $P$ and $R$ permutations and applies them directly without any explicit reordering. 
We store all data as AoS lexicographically.
Consequently, $P$ and $R$ on the reference configuration have a band structure (at least for lower orders).
However, this band remains relatively wide and sparse.


\subsection{Sparse Matrix Format}
To improve the arithmetic density of the matrix-vector multiplication, $P$ or $R$, have to be stored in an appropriate sparse format. The possible options for $P$ based on its structure are either the compressed sparse row (CSR) format, or the block compressed sparse row format (BSR). 
Column- or diagonal-based formats are not a fit, as we typically pick a subblock of $P$ or $R$ to be applied to one of the halo layers on the finer mesh.


We employ CSR despite the common knowledge that it challenges vectorisation \cite{SparseMatrices}.
We found that the matrix remains insufficiently band-structured and small, and direct BLAS or GEMM operations thus do not pay off once we switch to a block format.
The 58 unknowns of CCZ4 in return allow us to vectorise due to the AoS storage within the halo.


\section{Results}
\label{section:results}

\subsection{Accuracy}
\begin{figure*}
\centering
    \begin{subfigure}{0.33\textwidth}
        \includegraphics[width=0.9\textwidth]{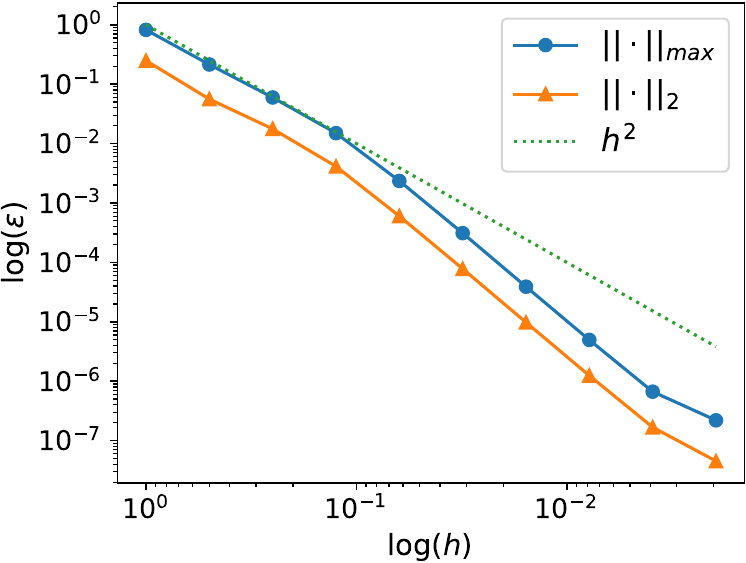}
        \caption{Matrix Interpolation}
    \end{subfigure}
    \hfill
    \begin{subfigure}{0.33\textwidth}
        \includegraphics[width=0.9\textwidth]{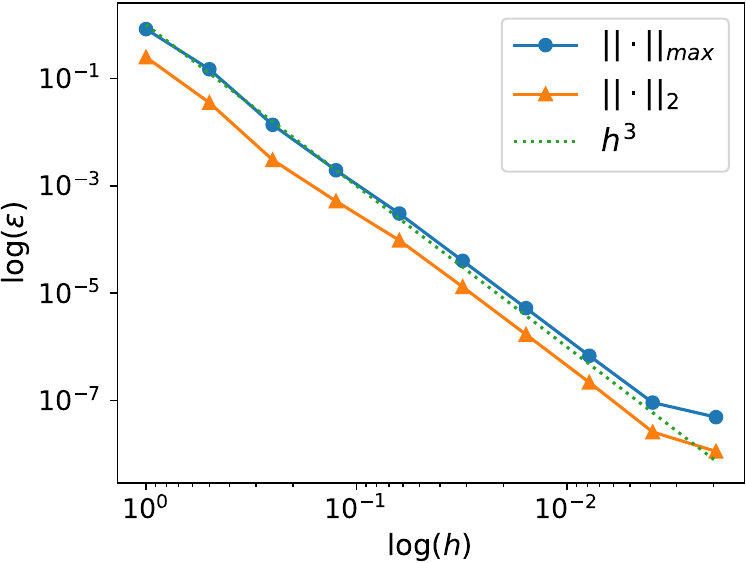}
        \caption{Second Order Interpolation}
    \end{subfigure}
    \hfill
    \begin{subfigure}{0.33\textwidth}
        \includegraphics[width=0.9\textwidth]{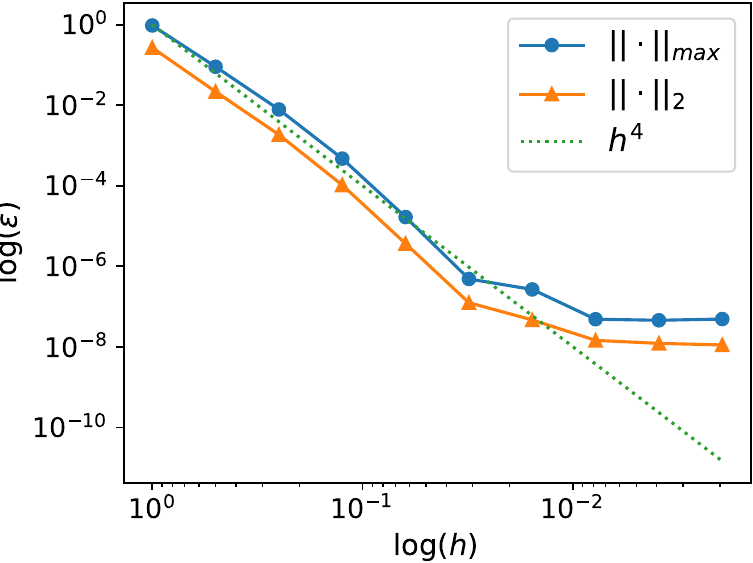}
        \caption{Third Order Interpolation}
    \end{subfigure}
    \caption{The error convergence plot for each interpolation scheme. The logarithm of the error norm, $\log(\epsilon)$, is plotted against $log(h)$. The order of convergence is obtained by comparing this to plots of $h^p$ where $p$ is a possible order of convergence.}
    \label{fig:convergence_interpolation}
\end{figure*}

To validate the convergence order of our numerical schemes, we take samples from an infinitely differentiable function $\sin(x)$ with non-vanishing derivatives on the coarse or fine data for the interpolation and the restriction respectively. In order to assess convergence, we reduce the spacing between fine cells $h$ while the ratio between coarse resolution and fine resolution is always 3:1 as we impose 2:1 balancing over our spacetree. For each value of $h$ used the interpolated or restricted values are compared to the analytic function. Two norms, $\lVert \cdot \rVert_{max}$ and $\lVert \cdot \rVert_{2}$, are used to assess the overall accuracy for the mesh resolution. The convergence plots for these experiments are shown in Figure \ref {fig:convergence_interpolation}.

The convergence order for ExaHyPE's now baseline matrix-based implementation is $O(h^2)$. This is as expected, as each function is approximated to be linear, leaving second order error terms. 
Similarly, the second- and third-order schemes show matching convergence according to their Taylor expansions. However, their convergence breaks down around an error of $10^{-8}$. This is a large enough error that it can not be attributed to the accuracy limitations of double precision variables. The most likely cause of this problem is an instability in the inversion of a triangular matrix within the QR decomposition. Simulations in this regime of mesh sizes are computationally infeasible for CCZ4 given its calibration regimes of interest. Hence, this break-down does not require further attention. 

The restriction shows analogous behaviour (not shown in this paper).

\subsection{Long-term behavior of stationary black holes}

We simulate black holes in the second-order CCZ4 formulation through the software ExaGRyPE which is built on top of Peano/ExaHyPE and employs the interpolation and restriction routines introduced in this paper.
A standard test case to validate any numerical relativity code is the simulation of a stationary black hole (cmp.~experimental description in \cite{ExaGRyPE}). 
For the present numerical tests, we use an adaptive Cartesian mesh refining around the black hole:

The domain is of size $[-9M,9M]^3$ and we refine the domain at radius $r = 7M$ and again at $r = 3M$.
The black hole is placed at
the origin of the three-dimensional domain, with an ADM
mass M = 1 and zero spin S = 0. 

Solutions to the Einstein field equations have to obey the Hamiltonian and Momentum constraints. As part of our studies, we post-process the violation of the Hamiltonian and the space-time lapse over a 1D cut through the domain.
Close to the black hole, a strong violation can be accepted.
Reasonably away from the singularity, the violation should be close to zero numerically.

\begin{figure}[H]
    \centering
    \begin{subfigure}{0.75\textwidth}
        \includegraphics[width=\textwidth]{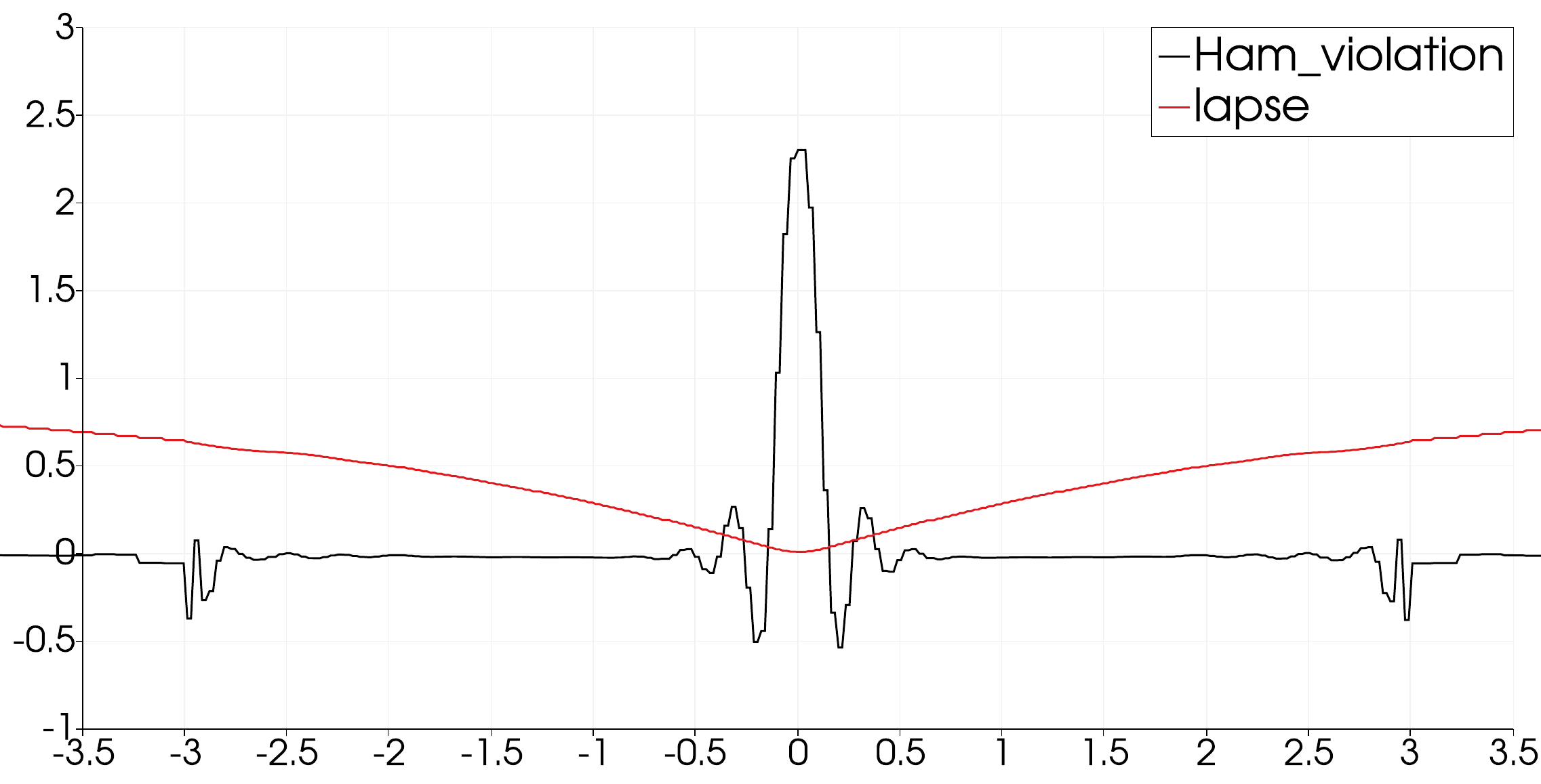}
        \caption{Tensor Product}
    \end{subfigure}
    \begin{subfigure}{0.75\textwidth}
        \includegraphics[width=\textwidth]{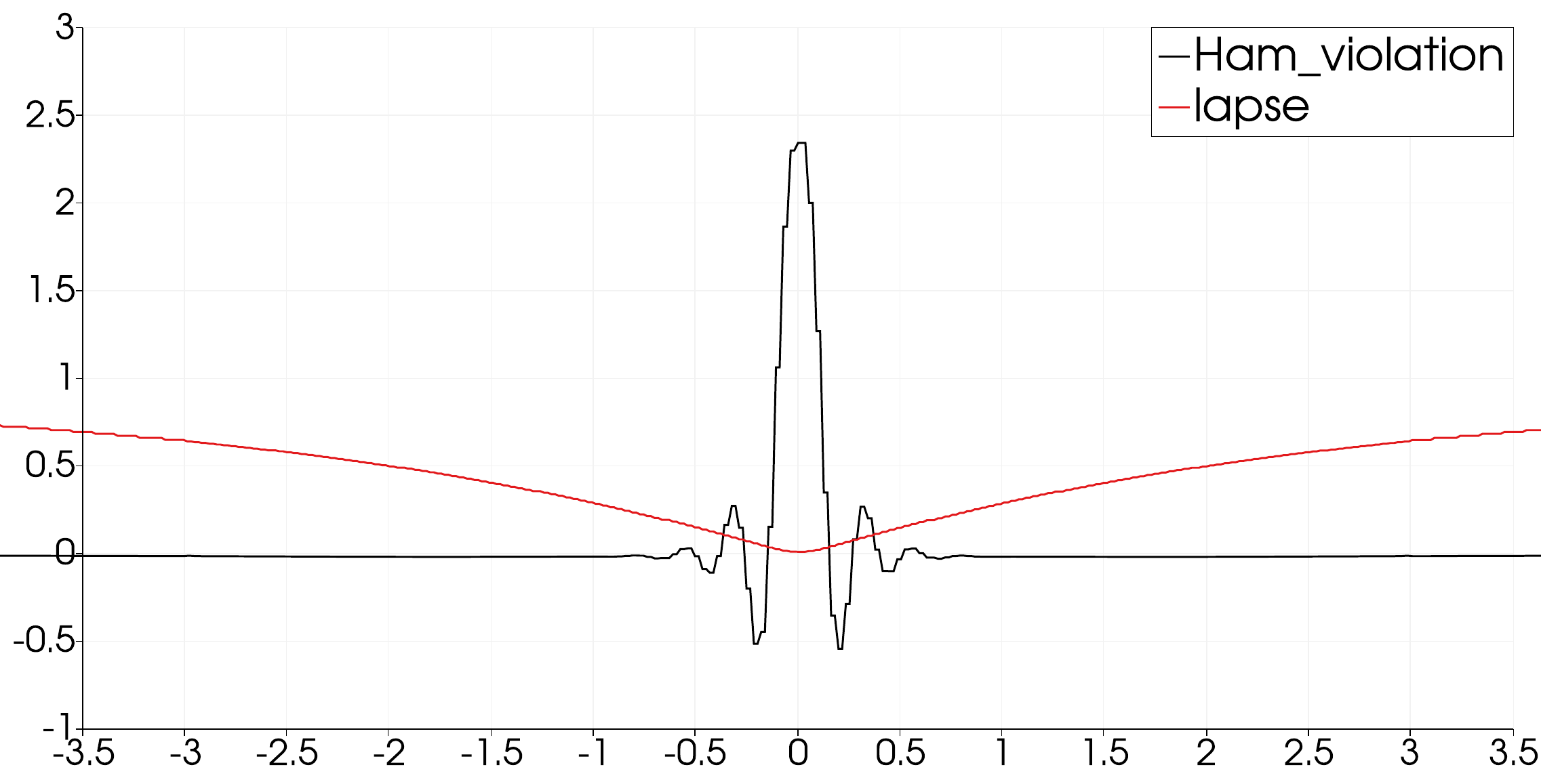}
        \caption{Second Order}
    \end{subfigure}
    \caption{
      Plots of the Hamiltonian constraint errors along the domain's x axis for a simulation of a stationary black hole using the tensor product and second order interpolations and restriction.
      \label{black_hole}
    }
\end{figure}

The violation of the Hamiltonian within the domain coincides spatially with the AMR mesh transition, at $x=\pm 2.5$ if we use trilinear interpolation (Figure \ref{black_hole}). 

Using a higher order interpolation and restriction can be seen to remove these errors.

Further measurements show that the violation's amplitude increases over time for first-order interpolation, indicating an accumulating error which eventually leads to an unstable simulation. As such, a higher-order interpolation help to resolve this issue, and leads to significantly improved adaptivity over longer simulation times, i.e.~we can simulate over longer time spans or use more aggressive spatial adaptivity. 
\subsection{Runtime Performance}

 All tests were executed on a
 dual 32-core AMD EPYC 7542 processors with a frequency of 2.9GHz. 
Each method was benchmarked individually by running it 100 times without reallocating any arrays.
We consequently eliminate memory allocation effects. 

\begin{figure}
\centering
    \begin{subfigure}{0.48\textwidth}
        \includegraphics[width=\textwidth]{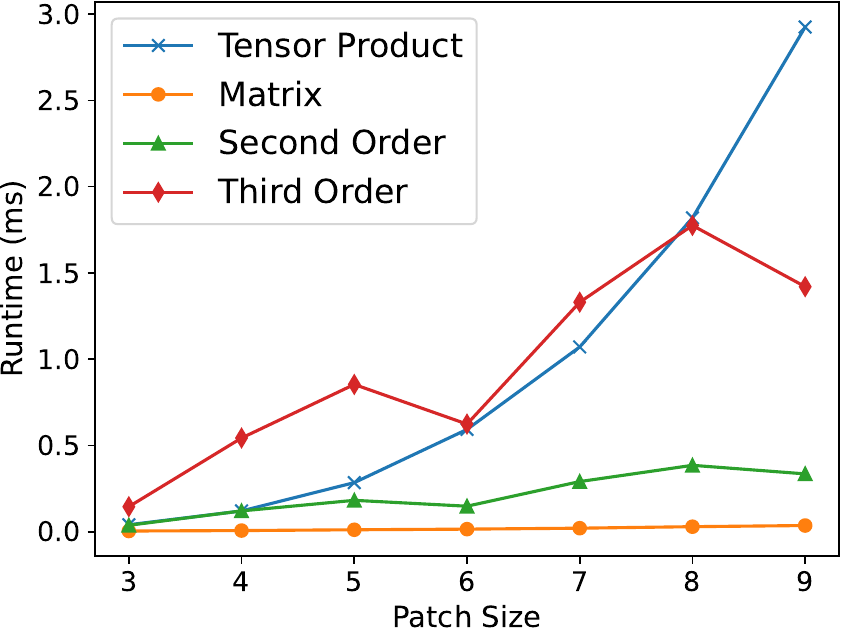}
        \caption{Run-times}
    \end{subfigure}
    \hfill
    \begin{subfigure}{0.48\textwidth}
        \includegraphics[width=\textwidth]{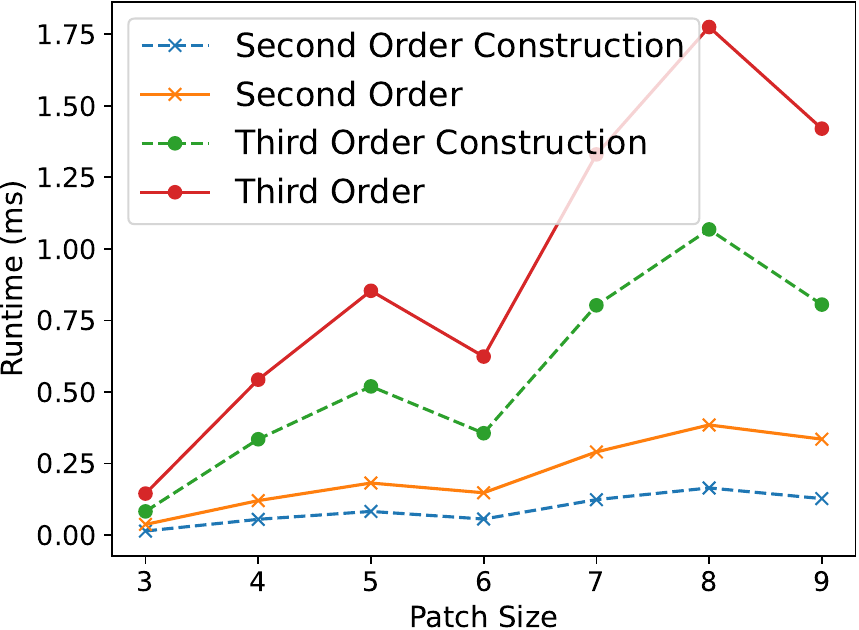}
        \caption{Matrix Construction}
        \label{fig:construction}
    \end{subfigure}
    \hfill
    
    \caption{
      Left: Runtime for each interpolation method against patch size. Right: Time spent in matrix construction for higher order interpolation compared to total run-time
      \label{fig:interpolation_times}
    }
\end{figure}

For each measurements the patch size is varied with the halo size fixed at 3 (Figure \ref{fig:interpolation_times}).
The restriction data follow the timings of the interpolation.

The vanilla tensor product implementation's runtime increases with the patch size $p$ quadratically, which matches the face's growth.
The low-order matrix version is almost immune to changes in $p$.

The higher order methods have worse scaling with the patch size since larger matrices have to be computed then inverted (Figure \ref{fig:construction}). Precomputing these matrices for each possible stencil configuration would therefore yield large performance gains. The methods perform slightly better when patch size is a multiple of 3 due to the preferable alignment of the coarse and fine cells.

In the stationary black hole benchmark, we stick to single node experiments to avoid network data exchange effects.
Consequently, our computational domain is tiny and lacks physically reasonable resolutions (cmp.~the violation of the Hamiltonian around the black hole which is a direct consequence of an insufficient mesh width). 
Larger meshes on multiple nodes will introduce a more pronounced adaptivity pattern---comparable literature mentions up to a resolution difference of 10--12---and hence make the imprint of interpolation and restriction more pronounced.

\section{Outlook and Conclusion}
\label{section:conclusion}

We introduce a higher-order interpolation and restriction scheme which replaces ExaHyPE's vanilla trilinear operators.
For a stationary black hole, the higher-order schemes eliminate the numerical inaccuracy along the AMR transitions that leads to constraint violations and long-term instabilities.
Tested in isolation, even our higher-order schemes outperform the baseline implementation (Figure \ref{fig:interpolation_times}).

As a shortcoming of the current matrix implementation, the CSR format does not allow for optimised Basic Linear Algebra Subprograms (BLAS) routines to be used, which limits the methods performance. Block or band matrices are worth considering as an alternative since they can use general matrix multiplication (GEMM) operations. For CCZ4 with 58 unknowns/PDEs to be evolved, BLAS is not critical. However, for ``simpler'' PDEs, resorting to a BLAS-based implementation might be crucial.

Baked into the current realisation is the lack of access to diagonal values  along the edges and vertices of a patch. The stencil construction accommodates this limitation, but a change to the underlying meshing logic would improve the method's accuracy. 
Many codes in the field omit diagonal entries as their management complicates and increases data movements and synchronisation over MPI boundaries significantly.
Performance-wisely, it is not clear whether adding diagonal data would improve the usability of the overall code base.

Our operators seem to be well-suited for a GPU port. However, such an endeavour is likely only successful if the underlying patch update is ported to GPUs, too, as the AMR boundary update is lightweight overall and would suffer from memory movements.
This is a subject of future work.

\bibliography{bibliography}

\appendix

\newpage
\section{Testing}

The different methods of interpolation and restriction are tested both individually, and as part of a larger application. It is assumed that the existing tensor product formulation gives accurate results for linear functions to simplify initial unit testing, in which the results of each new scheme are directly compared to the tensor product results. This is done for every possible face configuration, with all initial data generated using a linear function. This unit testing exists for the purpose of ensuring code correctness, and it results in simply a pass or fail.

The accuracy of these methods is further tested by performing convergence tests for each function. Coarse or fine data is generated using an infinitely differentiable function, then each method is used to interpolate and restrict from this data. The results of this are then compared to the original analytic function.

The final test for accuracy is to run a simulation of a stationary black hole that integrates these methods into the rest of the ExaHyPE code. This exists to check visually for Hamiltonian constraint violations near the AMR boundaries by creating snapshots of the simulation.

The speed of these methods is also tested both as a unit, and as an integrated system. Each function is run 100 times to create average runtimes, which are compared directly as the main metric for improvement throughout development. The \lstinline{performance_testbed} benchmark from the Peano codebase \cite{PeanoCode} is also used to measure the runtime of each function in order to determine how much time is spent on the interpolation and restriction stage of the simulation.
\section{Code Overview}
ExaHyPE uses C++ for its runtime code, with Python scripts used to generate some code. For the tensor product and matrix formulations, all matrices were generated and injected into the code at compile time using Python. The kernels used to update each patch are generated similarly \cite{ExaGRyPE}.

Peano uses an overall structure in which the most general code, which is useful for any problem, is part of Peano itself. There are new projects with their own code built from this, such as ExaHyPE. For this project the work is general, as Peano always uses an AMR approach, so the Peano code is modified. ExaHyPE is used in this project as a means to evaluate the new schemes. ExaHyPE code is only touched to modify specific solvers so that they have access to any generated matrices used in the interpolation or restriction.

Each interpolation is performed in a single function \\ \texttt{interpolateHaloLayer\_AoS\_<scheme>} which takes in parameters to express the dimensions and orientation of the face being interpolated on, the coarse data of the face in the form of an array, and an output array for the resulting fine data. A similar approach is used for the restriction method \texttt{restrictInnerHalfOfHaloLayer\_AoS\_<scheme>}, although this function needs to be called separately to restrict each half of the face. Since some of the schemes require additional arguments, such as matrices, secondary functions with fixed arguments are used to call them. These secondary functions access the specific solver being used to identify which matrices should be used, if any. This limits how much code needs to be modified when the scheme used is changed. 
\section{Stencil Structure}
\label{appendix:stencil}

\begin{figure}[h]
    \centering
    \includegraphics[width=0.8\textwidth]{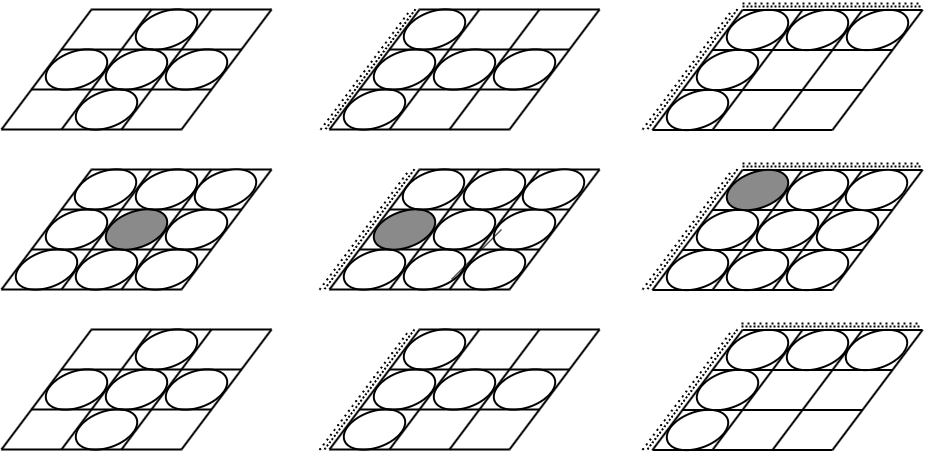}
    \caption{Three possibles arrangements of the coarse cells being considered for the second order interpolation. The shaded ellipse indicates the coarse cell containing the current fine cells being interpolated. Any dotted lines indicate the edge of a face. Through reflections and axis permutation, these three arrangements can be modified to encompass any possible scenario. Left: The default case in the centre of the face. Centre: The case where the centre coarse cell is against one face boundary. Right: The case where the centre coarse cell is against two face boundaries.}
    \label{fig:stencil_second}
\end{figure}

To determine the derivatives, we fit polynomials through the coarse cell centres. These polynomials are not unique: Within a patch, we can use a cell-centred stencil, or we can fit polynomials that are biased into one direction. This freedom helps us to construct the polynomials close to patch edges and vertices, as we have no ``diagonal'' halo patch data: We bias the polynomial fitting towards the centre of the preimage patch of the interpolation. We extrapolate the solution to interpolate. Within the cell centres, we favour central differences (Figure~\ref{fig:stencil_second}).

The third-order interpolation stencil is the same as that used by Colella and McCorquodale \cite{HigherOrderInterpolation}.

\end{document}